\pdfoutput=1
\documentclass[
  twocolumn,
  amsmath,amssymb,
  aps,
  superscriptaddress,
]{revtex4-2}

\usepackage{graphicx}
\usepackage{dcolumn}
\usepackage{bm}
\usepackage[hidelinks]{hyperref}
\hypersetup{
    colorlinks=true,
    linkcolor=blue,    
    citecolor=blue,    
    urlcolor=blue,     
}
\usepackage{array}
\usepackage{tabularx}
\usepackage[dvipsnames]{xcolor}
\usepackage{amsmath}
\DeclareMathOperator*{\argmax}{arg\,max}
\usepackage{multirow}
\usepackage{booktabs}
\usepackage{longtable}
\usepackage{float}
\makeatletter
\let\float@end\float@end@ltx
\let\newfloat\newfloat@ltx
\let\float@end@float\float@end@ltx
\let\newfloat@float\newfloat@ltx
\makeatother
\usepackage{makecell}

\usepackage{algorithm}%
\usepackage{algorithmicx}%
\usepackage{algpseudocode}%
\makeatletter
\renewcommand\fnum@algorithm{\fname@algorithm\ \thealgorithm}
\makeatother

\usepackage{setspace}
\usepackage{etoolbox}
\usepackage{listings}
\definecolor{codebg}{RGB}{247,247,247}
\definecolor{codegreen}{RGB}{0,128,0}
\definecolor{codeblue}{RGB}{0,0,180}
\definecolor{codepurple}{RGB}{160,32,160}
\definecolor{codegray}{RGB}{100,100,100}
\AtBeginEnvironment{lstlisting}{\singlespacing}

\begin{document}

\title{Qmes: Quantum Meta-Learning for Encoding Selection in Quantum Kernel Methods}

\author{Dao Duy Tung}
\affiliation{Faculty of Physics and Engineering Physics, University of Science, Ho Chi Minh City 70000, Vietnam}
\affiliation{Vietnam National University, Ho Chi Minh City 70000, Vietnam}

\author{Quoc Chuong Nguyen}
\email[Corresponding author: ]{nguyenquocchuong2@duytan.edu.vn}
\affiliation{Institute of Fundamental and Applied Sciences, Duy Tan University, Ho Chi Minh City 70000, Vietnam}

\author{Vu Tuan Hai}
\email[Corresponding author: ]{haivt@uit.edu.vn}
\affiliation{University of Information Technology, Ho Chi Minh City 70000, Vietnam}
\affiliation{Vietnam National University, Ho Chi Minh City 70000, Vietnam}

\author{Le Bin Ho}
\email[Corresponding author: ]{ho.bin.le.e3@tohoku.ac.jp}
\affiliation{Graduate School of Engineering, Tohoku University, Sendai 980-8579, Japan}
\affiliation{Frontier Research Institute for Interdisciplinary Sciences, Tohoku University, Sendai 980-8578, Japan}

\author{Lan Nguyen Tran}
\email[Corresponding author: ]{tnlan@hcmus.edu.vn}
\affiliation{Faculty of Physics and Engineering Physics, University of Science, Ho Chi Minh City 70000, Vietnam}
\affiliation{Vietnam National University, Ho Chi Minh City 70000, Vietnam}


\begin{abstract}
Selecting an effective encoding quantum circuit is a key challenge in quantum kernel methods because different feature maps can lead to different performance. Conventional methods require constructing and evaluating every circuit for each new dataset, making it computationally expensive. We present Qmes, an open-source Python package that automatically recommends circuits through meta-learning. Qmes characterizes a dataset using classical complexity measures and queries a pre-trained model to recommend circuits without quantum evaluation at inference time. The package provides modular components for meta-feature extraction, quantum-kernel evaluation, recommender training, model selection, and user-defined circuit extension. We validate Qmes on 105 classification and 86 regression benchmark datasets. Qmes reduces the mean recommendation regret by 2.2$\times$ and 4.2$\times$ for classification and regression, respectively, compared to a non-adaptive baseline, with statistical significance confirmed via a paired Wilcoxon signed-rank test ($p<10^{-4})$. Qmes thus enables efficient and practical encoding-circuit selection for quantum kernel methods.
\end{abstract}


\maketitle

\noindent
\textbf{PROGRAM SUMMARY}\\
\noindent
\textit{Program title:} Qmes \\
\textit{CPC Library link to program files:} (to be added by Technical Editor) \\
\textit{Developer's repository link:} \url{https://github.com/tungduy1704/Qmes},
  \url{https://tungduy1704.github.io/Qmes} \\
\textit{Code Ocean capsule:} (to be added by Technical Editor) \\
\textit{Licensing provisions:} MIT License  \\
\textit{Programming language:} Python 3 (requires Python $\geq$ 3.10; tested on 3.10--3.12) \\
\textit{External routines/libraries:} Problexity; Qsun simulator; NumPy, pandas, scikit-learn  \\
\textit{Nature of problem:}
In quantum kernel methods, predictive performance depends strongly on the choice of encoding circuit. Selecting an effective circuit for a given dataset conventionally requires an exhaustive search from multiple candidate circuits, making the selection process computationally expensive.\\
\textit{Solution method:}
Qmes formulates encoding-circuit selection as a supervised meta-learning problem with input is the data complexity extracted from the dataset and output is the recommended circuit. We create a model that can be trained on this meta-dataset; then the trained model can perform inference on new datasets. \\
\textit{Additional comments including Restrictions and Unusual features:}
The quantum cost is incurred once during meta-dataset construction, and the resulting model is distributed in serialized form, making the inference process entirely classical. Restrictions: the package targets supervised tabular classification and regression, since the complexity measures require a labelled target. Circuit evaluation uses noiseless state-vector simulation; hardware execution and noise models are not supported.


\section{\label{sec:intro}Introduction}

Early works in quantum machine learning (QML) extended standard data-analysis methods to quantum settings, including principal component analysis~\cite{Lloyd2014}, support vector machines~\cite{PhysRevLett.113.130503}, and topological data analysis~\cite{Lloyd2016}.
These algorithms promise large asymptotic speedups, but they presuppose coherent access to the classical input through quantum random access memory (QRAM) and structural conditions such as low rank, neither of which can be taken for granted in practice. Attention has since shifted to models that encode the classical input directly into a quantum state through a parameterized circuit~\cite{lloyd2020quantumembeddingsmachinelearning}. Such classifiers have been implemented on superconducting processors~\cite{Havlicek2019} and applied to the cosmological benchmark of supernova classification from real spectral features~\cite{Peters2021}. Across these methods, the classical input is mapped to a quantum state in a Hilbert space, and learning proceeds through that map \cite{schuld2021supervisedquantummachinelearning}. Two families of models are built on it: quantum neural networks, which train gates applied after the map \cite{PhysRevA.98.032309}, and quantum kernel methods (QKMs), which keep the map fixed and delegate optimization to a classical kernel algorithm \cite{PhysRevLett.122.040504}.

In a QKM, this feature map is used to define a kernel as the inner product between the resulting quantum states, providing a measure of how similar two inputs appear once embedded in the Hilbert space. Because kernel-based learning algorithms operate entirely on these pairwise similarity values rather than on the feature vectors themselves, the encoding circuit is not a neutral preprocessing step: it determines the kernel, and therefore the notion of similarity on which the learning algorithm relies. Different circuits induce different kernels, and hence different hypothesis classes, even when the same classical algorithm is applied downstream. The circuit is thus a key design choice that can strongly influence the model performance \cite{schuld2021supervisedquantummachinelearning}.

The design question is how to identify the most effective circuit for a given dataset. One line of work answers this question by exhaustive evaluation: each circuit is scored on the target dataset, and the best-performing one is retained \cite{Jha2026}. However, this approach scales poorly. A single evaluation requires constructing kernel matrix with $\mathcal{O}(n^2)$ pairwise kernel evaluations for $n$ samples. Neither the kernel matrix nor the resulting ranking can be reused between different datasets. Therefore, the procedure must be repeated for every circuit on every dataset encountered.

A second line of work seeks to automate this choice. However, existing approaches do not fully address the requirements of the present setting. Incudini \textit{et al.}~\cite{10812182} discovered circuits by combinatorial optimization, but their method requires a new optimization loop for every dataset rather than learning from dataset characteristics. Neto \textit{et al.}~\cite{neto2025datacomplexitymeasuresquantum} learned from such characteristics using classical complexity measures, but their approach targets variational circuits rather than quantum kernels. In our previous work~\cite{tung2026automatedselectionquantumencoding}, we first demonstrated that dataset-aware selection is feasible for quantum kernels by recommending circuits for unseen classification datasets using complexity metrics alone. However, that study was limited to a proof-of-concept for classification and did not provide a documented software package, pre-trained models, or support for regression tasks.

We present Qmes, an open-source Python package that implements the pipeline in Ref.~\cite{tung2026automatedselectionquantumencoding}. Beyond classification, Qmes extends the method to regression problems and adopts a one-vs-one (OvO) pairwise formulation. The trained recommender is saved as a pre-trained model and can be reused to recommend circuits for new datasets without additional quantum-kernel evaluation.
\section{Method and Software Description}\label{sec:description}
\subsection{Problem Formulation}

Quantum kernel models extend classical kernel learning to the quantum setting by encoding classical inputs as quantum states $|\phi(\boldsymbol{x})\rangle = U(\boldsymbol{x})|0\rangle^{\otimes n_q}$, where $U(\boldsymbol{x})$ is a circuit that acts as a feature map in an $n_q$-qubit Hilbert space. The quantum kernel is then defined as
\begin{equation}
  K(\bm{x}_j, \bm{x}_i) =
  |\langle\phi(\bm{x}_j)|\phi(\bm{x}_i)\rangle|^2.
\end{equation}
Since the kernel defines a similarity measure, $U$ determines the geometry in which this similarity is computed \cite{Huang2021}. An effective $U$ arranges the data in this geometry such that samples considered similar by the learning task lie close together, while dissimilar samples lie farther apart~\cite{lloyd2020quantumembeddingsmachinelearning}. Identifying a well-matched $U$, however, is costly and must be repeated for every new dataset. To reduce this cost, a proxy can be used to predict circuit performance without evaluating every candidate circuit. Here, the proxy takes the data complexity metrics as input, which characterize the dataset independently of any specific learning algorithm \cite{10.1145/3347711, 990132}. Prior work has shown that such metrics can serve as meta-features to predict which classical classifier is most suitable for a given dataset \cite{22}, and more recently, to guide the selection of quantum circuit architectures \cite{neto2025datacomplexitymeasuresquantum, tung2026automatedselectionquantumencoding}.

\begin{figure}[t]
  \centering
  \includegraphics[width=\linewidth]{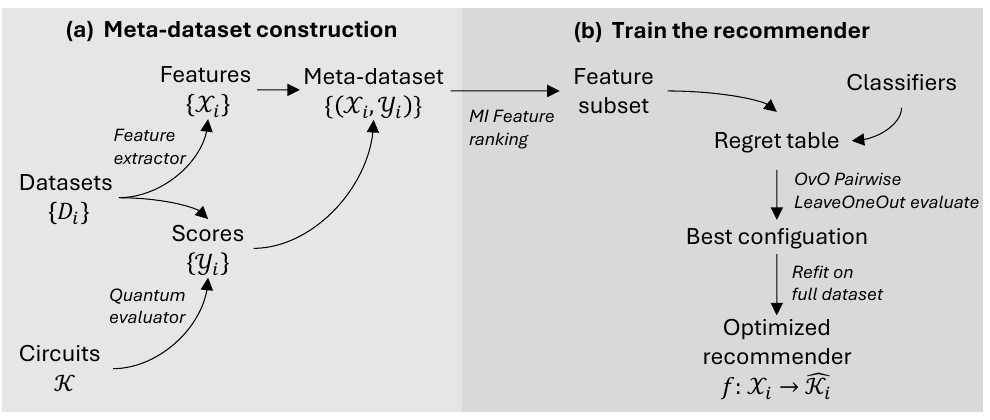}
  \caption{Overview of the proposed circuit selection framework. Stage (a): data processing and ground-truth label generation. Stage (b): recommender training with multiple classifiers and feature subset configurations.}
  \label{fig:workflow}
\end{figure}

Qmes formalizes this selection task as a supervised meta-learning problem. The core components follow the two-stage pipeline shown in Fig.~\ref{fig:workflow}. In the construction phase, all circuits in the pool $\mathcal{K}$ are evaluated on a collection of datasets $D$ using a kernel evaluator. Each dataset $\mathcal{D}_i$ is described by a complexity meta-feature vector $\mathcal{X}_i \in \mathbb{R}^d$ and a circuit performance score vector $\mathcal{Y}_i \in \mathbb{R}^{\lvert\mathcal{K}\rvert}$, forming a meta-dataset $\mathcal{D}_{\text{meta}} = \{(\mathcal{X}_i, \mathcal{Y}_i)\}_{i=1}^N$. A classical machine learning model (recommender) $f$ is then trained on $\mathcal{D}_{\text{meta}}$ to learn the mapping $\mathcal{X}_i \mapsto \mathcal{Y}_i$. At the inference stage, a new dataset is characterized by its complexity vector $\mathcal{X}_k$, and $f$ produces a ranked list of recommended circuits without additional quantum evaluation.

\subsection{Architecture Overview}

\begin{figure*}[t]
  \centering
  \includegraphics[width=0.8\linewidth]{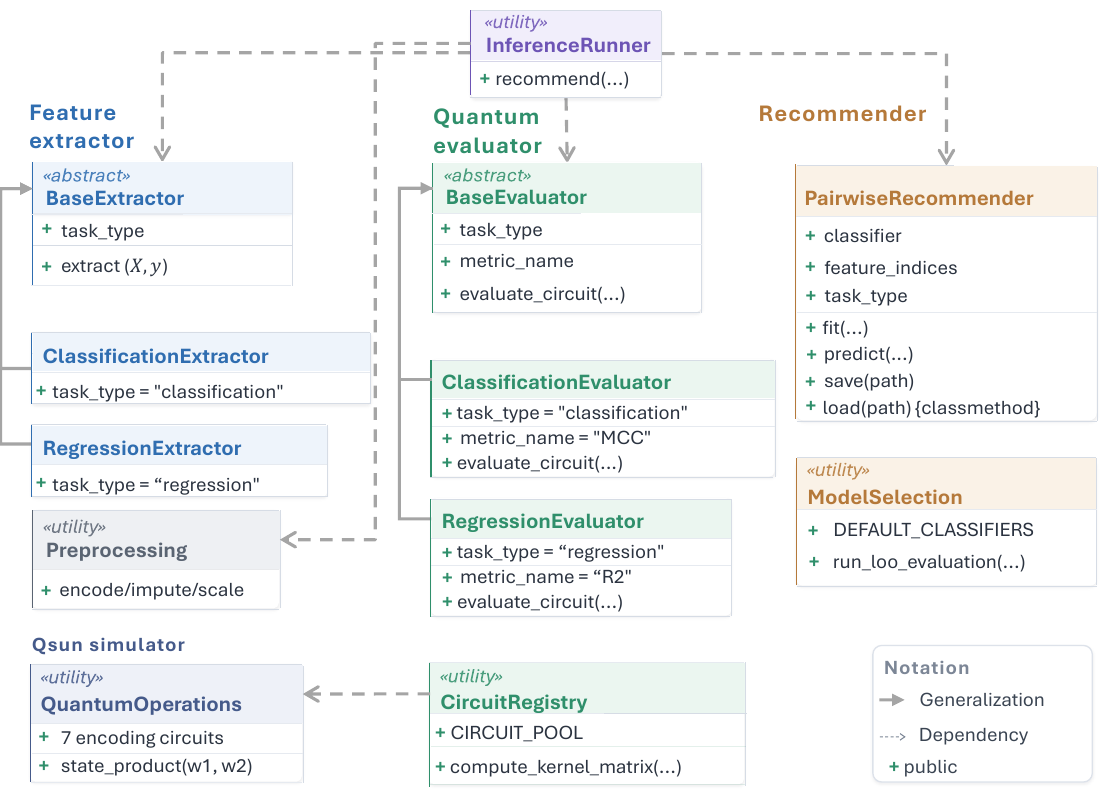}
  \caption{Software architecture of Qmes. The three \emph{core components} are defined as abstract base classes with task-specific subclasses, and are supported by four \emph{utility modules}.
  }
  \label{fig:architecture}
\end{figure*}

The Qmes architecture is illustrated in Fig.~\ref{fig:architecture}. The three core components are \texttt{FeatureExtractor}, \texttt{QuantumEvaluator}, and \texttt{Recommender}.
\begin{itemize}
    \item In the \texttt{FeatureExtractor} component, the class \texttt{BaseExtractor} and its subclasses \texttt{ClassificationExtractor} and \texttt{RegressionExtractor} implement stage~(a) of Fig.~\ref{fig:workflow}: given a raw dataset, they extract a complexity vector and return it in an \texttt{ExtractionResult} (a dataclass bundling the vector, its feature names, and task type).
    \item In the \texttt{QuantumEvaluator} component, the class \texttt{BaseEvaluator} and its subclasses \texttt{ClassificationEvaluator} and \texttt{RegressionEvaluator} complete stage~(a) by evaluating each circuit as a quantum kernel and recording its performance score, thereby producing the vector $\mathcal{Y}_i$.
    \item In the \texttt{Recommender} component, the class \texttt{PairwiseRecommender} implements stage (b) of Fig.~\ref{fig:workflow}: it is trained offline on the assembled meta-dataset, then at inference time accepts a complexity vector and returns a ranked list of circuits via \texttt{predict()}.
\end{itemize}

Four utility modules support these core components. The \texttt{Preprocessing} handles raw inputs, categorical encoding, imputation, and feature scaling before they reach an Extractor or Evaluator. The \texttt{CircuitRegistry} stores the circuit pool $\mathcal{K}$ and provides kernel-matrix computation to the Evaluator. Both modules support stage~(a). The \texttt{ModelSelection} produces the fitted recommender for stage~(b) by searching over classifier and feature-subset configurations. The \texttt{InferenceRunner} provides the entry points \texttt{recommend} and \texttt{preprocess\_new\_dataset} which connect the components during inference. The Qsun simulator \cite{Nguyen_2022} serves as the quantum backend for all circuit evaluation, and is invoked exclusively during the offline phase; inference requires no quantum computation. Within Qsun, \texttt{QuantumOperations} implements the seven encoding circuits and the state-overlap primitive that \texttt{CircuitRegistry} calls to compute kernel matrices.

To extend Qmes with a new meta-feature source, users need only implement the three members shown in Listing~\ref{lst:extractor}.

\begin{lstlisting}[
    language=Python,
    caption={Adding a new meta-feature source by subclassing \texttt{BaseExtractor}.},
    label={lst:extractor}
]
from Qmes.extractors import BaseExtractor
class MyExtractor(BaseExtractor):
    @property
    def task_type(self) -> str:
        return "classification"
    @property
    def _feature_names(self) -> list[str]:
        return ["my_feature_1", "my_feature_2"]
    def _extract_raw(self, X, y=None):
        return [0.1, 0.2]
\end{lstlisting}

A new evaluator for scoring circuit performance on the meta-dataset can therefore be added without modifying \texttt{PairwiseRecommender} or the \texttt{InferenceRunner}.

\subsection{Core Components}
\subsubsection{\label{extractor} Extractors}
Before extraction, the dataset is preprocessed: missing values are median-imputed, and datasets with more than 600 samples are subsampled using stratified sampling for classification and random sampling for regression. Features are then min-max scaled to $[0,1]$ internally, as required by Problexity \cite{10.1145/3347711, komorniczak2023problexity}.

Both \texttt{ClassificationExtractor} and \texttt{RegressionExtractor} return a fixed-length vector of the complexity measures introduced by Lorena et al. \cite{10.1145/3347711}, computed on the preprocessed data: 22 dimensions for classification and 12 for regression. For regression, the target $y$ is additionally scaled to $[0,1]$ before extraction. Two measures for each task are stochastic: for classification, $l_3$
(linearity) and $n_4$ (neighborhood); for regression, $l_3$ (linearity) and
$s_4$ (smoothness). Full definitions are given in Appendix~\ref{appendix:features}. Qmes averages each measure over 10 fixed seeds (0-9) to obtain deterministic outputs. Collected across all datasets, these vectors form the meta-feature matrix used for recommendation (Table~\ref{tab:meta_dataset}).

\subsubsection{\label{evaluator} Evaluators}
The Evaluator measures the performance of each circuit on a benchmark dataset, producing the scores that serve as labels for the meta-dataset. Both \texttt{ClassificationEvaluator} and \texttt{RegressionEvaluator} use a fixed quantum kernel and a fixed cross-validation random state, so scores are deterministic and reflect only the feature map contribution.

Both evaluators score each circuit using $3$-fold cross-validation with a precomputed kernel, stratified for classification and standard for regression. Within each fold $\ell \in \{1,2,3\}$, the features are preprocessed by using \texttt{StandardScaler}, reduced by Principal Component Analysis (PCA) to at most $4$ components to match the number of qubit, and then rescaled by using \texttt{MinMaxScaler} to $[0,\pi]$ (or $[0,1]$ for \texttt{unit}). All transformations are fitted on the training split only. Cross-validation is used instead of a single train/test split to reduce the variance of $\mathcal{Y}_{i, k}$ against any one particular partition, especially for the smaller benchmark datasets. Using 3 folds specifically is a practical design choice, given the cost of quantum-kernel evaluation.

For classification, circuit $c_k$'s predictions on the held-out fold $\ell$ of dataset $\mathcal{D}_i$ are produced by a Support Vector Classifier (SVC) \cite{Cortes1995} trained on the quantum kernel, and scored with the Matthews correlation coefficient~(MCC) \cite{10.1371/journal.pone.0177678}:
\begin{widetext}
    \begin{equation}
  \text{MCC}_{i,k}^{(\ell)} = \frac{TP \cdot TN - FP \cdot FN}
  {\sqrt{(TP+FP)(TP+FN)(TN+FP)(TN+FN)}} \in [-1, 1],
  \label{eq:mcc}
\end{equation}
\end{widetext}
where $TP$, $TN$, $FP$, and $FN$ represent true positives, true negatives, false positives, and false negatives on fold $\ell$. MCC equals $+1$ for perfect classification, $0$ for a classifier no better than random guessing, and $-1$ when every prediction is inverted relative to the true label. Averaging Eq.~\ref{eq:mcc} across the $3$ folds gives the corresponding entry of the circuit-score vector $\mathcal{Y}_i$ introduced in Sec.~\ref{sec:description}:
\begin{equation}
  \mathcal{Y}_{i,k}^{(MCC)} = \frac{1}{3}\sum_{\ell=1}^{3} \text{MCC}_{i,k}^{(\ell)}.
  \label{eq:mcc-avg}
\end{equation}

For regression, circuit $c_k$'s predictions on the held-out fold $\ell$ of dataset $\mathcal{D}_i$ are produced by Kernel Ridge Regression (KRR) \cite{10.5555/645527.657464} trained on the quantum kernel, with the target standardized within the fold and inverse-transformed before scoring by the coefficient of determination:
\begin{equation}
  R^{2\,(\ell)}_{i,k} = 1 - \frac{\sum_j (y_j - \hat{y}_j)^2}{\sum_j (y_j - \bar{y})^2} \in (-\infty, 1],
  \label{eq:r2}
\end{equation}
where $y_j$ and $\hat{y}_j$ denote the true and predicted target values for sample $j$ in fold $\ell$, and $\bar{y}$ is their mean over the fold. $R^2$ equals $1$ for perfect prediction, $0$ for a model no better than always predicting $\bar{y}$, and a value below $0$ on a held-out fold when the model performs worse than this baseline - unlike ordinary least-squares regression on training data, where $R^2$ is guaranteed non-negative. It measures the proportion of variance explained and is scale-invariant across datasets with different target ranges. Averaging Eq.~\ref{eq:r2} across the $3$ folds gives
\begin{equation}
  \mathcal{Y}_{i,k}^{(R^2)} = \frac{1}{3}\sum_{\ell=1}^{3} R^{2\,(\ell)}_{i,k}.
  \label{eq:r2-avg}
\end{equation}

Before the meta-dataset $\mathcal{D}_{\text{meta}}$ is assembled, each dataset $\mathcal{D}_i$ is checked against its circuit-score vector $\mathcal{Y}_i$ (Eq.~\ref{eq:mcc-avg} for classification, Eq.~\ref{eq:r2-avg}
for regression). $\mathcal{D}_i$ is removed as \emph{no-signal} if $\max_k \mathcal{Y}_{i,k} < 0.1$ (no circuit in the pool exceeds near-random performance), or as \emph{ceiling} if $\min_k \mathcal{Y}_{i,k} \ge 0.99$ (every circuit in the pool saturates,
leaving no discriminative signal to learn from). This yields the $N=105$ and $N=86$ datasets reported throughout, and guarantees $\max_k \mathcal{Y}_{i,k} \ge 0.1$ for every dataset entering the regret evaluation of Sec.~\ref{selection}.

\begin{table*}[t]
  \caption{Example rows from the classification meta-dataset $\mathcal{D}_{\text{meta}} = \{(\mathcal{X}_i, \mathcal{Y}_i^{(MCC)})\}_{i=1}^{N}$, showing 5 of the $N=105$ classification benchmark datasets (full list in Appendix~\ref{appendix:datasets}). Each column reports the complexity meta-feature vector $\mathcal{X}_i \in \mathbb{R}^{22}$ (6 of its 22 entries shown) and the circuit-score vector $\mathcal{Y}_i^{(MCC)} \in \mathbb{R}^{7}$, one mean-MCC entry per circuit in the pool (all 7 shown; see Appendix~\ref{circuit}). Regression follows the same schema on a separate meta-dataset $\mathcal{D}_{\text{meta}} = \{(\mathcal{X}_i, \mathcal{Y}_i^{(R^2)})\}_{i=1}^{N}$ of $N=86$ benchmark datasets: a 12-dimensional meta-feature vector $\mathcal{X}_i \in \mathbb{R}^{12}$ in place of the 22-dimensional vector shown here, and a circuit-score vector $\mathcal{Y}_i^{(R^2)} \in \mathbb{R}^{7}$, one mean-$R^2$ entry per circuit.
  }
  \label{tab:meta_dataset}
  \begin{tabular*}{0.8\linewidth}{@{\extracolsep{\fill}}lccccccc ccccccc@{}}
    \toprule
    & \multicolumn{7}{c}{\textbf{Meta-features} $\mathcal{X}_i$}
    & \multicolumn{7}{c}{\textbf{Circuit scores} $\mathcal{Y}_i^{(MCC)}$} \\
    \cmidrule(lr){2-8} \cmidrule(lr){9-15}
        \textbf{Dataset} & $f_1$ & $f_3$ & $n_2$ & $t_1$ & $\mathrm{lsc}$ & $c_2$ & $\cdots$
    & \texttt{unit} & \texttt{SRx} & \texttt{RY} & \texttt{HERx} & \texttt{RY\_CX} & \texttt{ZFM} & \texttt{HD} \\
    \midrule
    Iris\_01   & 0.03 & 0.00 & 0.09 & 0.03 & 0.51 & 0.00 & $\cdots$
               & 1.00 & 0.98 & 1.00 & 1.00 & 1.00 & 0.86 & 0.98 \\
    Wine\_01   & 0.17 & 0.21 & 0.35 & 0.36 & 0.78 & 0.02 & $\cdots$
               & 0.95 & 0.88 & 0.95 & 0.92 & 0.91 & 0.50 & 0.94 \\
    Breast-w   & 0.15 & 0.86 & 0.17 & 0.15 & 0.64 & 0.18 & $\cdots$
               & 0.94 & 0.94 & 0.95 & 0.95 & 0.94 & 0.90 & 0.94 \\
    Ionosphere & 0.54 & 0.80 & 0.19 & 0.31 & 0.90 & 0.15 & $\cdots$
               & 0.75 & 0.80 & 0.73 & 0.78 & 0.79 & 0.72 & 0.78 \\
    Moons      & 0.38 & 0.61 & 0.26 & 0.29 & 0.96 & 0.00 & $\cdots$
               & 0.69 & 0.61 & 0.68 & 0.71 & 0.60 & 0.21 & 0.71 \\
    $\vdots$   & $\vdots$ & $\vdots$ & $\vdots$ & $\vdots$ & $\vdots$ & $\vdots$ & $\vdots$
               & $\vdots$ & $\vdots$ & $\vdots$ & $\vdots$ & $\vdots$ & $\vdots$ & $\vdots$ \\
    \bottomrule
  \end{tabular*}
\end{table*}

\subsubsection{\label{pairwise}Pairwise OvO Recommender}

\texttt{PairwiseRecommender} is shared by both tasks and is trained offline on the meta-dataset $\mathcal{D}_{\text{meta}}$ constructed by the Extractor and Evaluator (Table~\ref{tab:meta_dataset}).

Rather than training a single classifier to select the optimal circuit, the Recommender decomposes the selection into pairwise comparisons. For each pair of circuits $(c_a,c_b)$, it trains a binary classifier, referred to as a \emph{comparator}. During training, the binary label for dataset $\mathcal{D}_i$ is determined by comparing the Evaluator scores of the two circuits:
\begin{equation}
  z_i^{(a,b)} =
  \begin{cases}
    1 & \text{if } \mathcal{Y}_{i,a} \geq \mathcal{Y}_{i,b} \\
    0 & \text{otherwise,}
  \end{cases}
  \label{eq:pairlabel}
\end{equation}
where $\mathcal{Y}_{i,k}$ denotes the score of the circuit $c_k \in \mathcal{K}$ on dataset $\mathcal{D}_i$. Each comparator thus learns which structural properties of a dataset favor $c_a$ over $c_b$, independently of the other pairs. At the inference stage, the complexity vector $\mathcal{X}_{k}$ is passed through all $\binom{|\mathcal{K}|}{2}$ comparators, each casts one vote for the circuit it predicts to be superior. The circuits are ranked according to their total vote count $\{v_k\}$, as shown in Algorithm~\ref{alg:ovo}.

\begin{algorithm}[t]
    \caption{OvO pairwise recommender: training and inference}
    \label{alg:ovo}
        
        \begin{algorithmic}
        \Require Meta-features $\mathcal{X}\in\mathbb{R}^{N\times d}$, scores $\mathcal{Y}\in\mathbb{R}^{N\times\lvert\mathcal{K}\rvert}$, circuit pool $\mathcal{K}$, base classifier $h$, query $\mathcal{X}_{k}$
        \Ensure Ranking and votes $\{v_k\}$ for $\mathcal{X}_{k}$
        \Statex \textit{// Training stage}
        \For{each pair $(c_a,c_b)$ with $\{(a,b)|1\leq a < b \le \lvert\mathcal{K}\rvert\}$}
            \State construct labels $\{z_i^{(a,b)}\}_{i=1}^{N}$ \Comment{see Equation~\ref{eq:pairlabel}}
            \State $h_{ab} \gets$ $\mathrm{Train}
            \left(h,\mathcal{X},\{z_i^{(a,b)}\}_{i=1}^{N}\right)$
        \EndFor
        \Statex \textit{// Inference stage}
        \State $v_k \gets 0$ for $k = 1,\dots,\lvert\mathcal{K}\rvert$
        \For{each pair $(c_a,c_b)$ with $\{(a,b)|1 \le a < b \le \lvert\mathcal{K}\rvert\}$}
            \If{$h_{ab}(\mathcal{X}_{k}) = 1$}
                \State $v_a \gets v_a +\ 1$
            \Else
                \State $v_b \gets v_b +\ 1$
            \EndIf
        \EndFor
        \State $\text{ranking} \gets \text{Sort}(\mathcal{K}$ by $v_k$ descending$)$

        \State \Return $\text{ranking},\ \{v_k\}_{k=1}^{\lvert\mathcal{K}\rvert}$
    \end{algorithmic}
\end{algorithm}

OvO is preferred over a direct $\lvert\mathcal{K}\rvert$-class formulation for two reasons. First, the meta-dataset is small ($N \approx 100$ datasets per task), and a direct $\lvert\mathcal{K}\rvert$-class classifier would have to learn $\lvert\mathcal{K}\rvert$ classes from this limited number of samples. In contrast, each classifier can use all $N$ datasets to learn a single, simpler pairwise boundary. Second, rather than collapsing the full score vector into a single best-circuit label, OvO retains the complete pairwise ordering information across all circuits, allowing finer distinctions when performance differences are small.

\subsection{\label{selection} ModelSelection Module}

The complexity features in $\mathcal{D}_{\text{meta}}$ differ in their relevance to the performance of the circuit. To identify the most informative features, each meta-feature is scored by mutual information~(MI) \cite{PhysRevE.69.066138, 10.1371/journal.pone.0087357} between its values across the datasets, $\{\mathcal{X}_{i,j}\}_{i=1}^{N}$. The corresponding best-circuit labels are $\{k^*_i\}_{i=1}^{N}$, where $k^*_i = \argmax_{k}\, \mathcal{Y}_{i,k}$. The features are ranked by MI in descending order. The best-circuit label $k_i^*$ is used only for feature ranking and does not replace the pairwise labels used by the OvO recommender. Subsets containing the top-$m$ ranked features are then evaluated alongside the full feature set, with $m \in \{5, 10, 15, 20, \text{full}\}$ for classification and $m \in \{5, 8, 10, \text{full}\}$ for regression. These values are a practically chosen grid, not a full search over every $m$.

\begin{figure*}[t]
  \centering
  \includegraphics[width=0.7\linewidth]{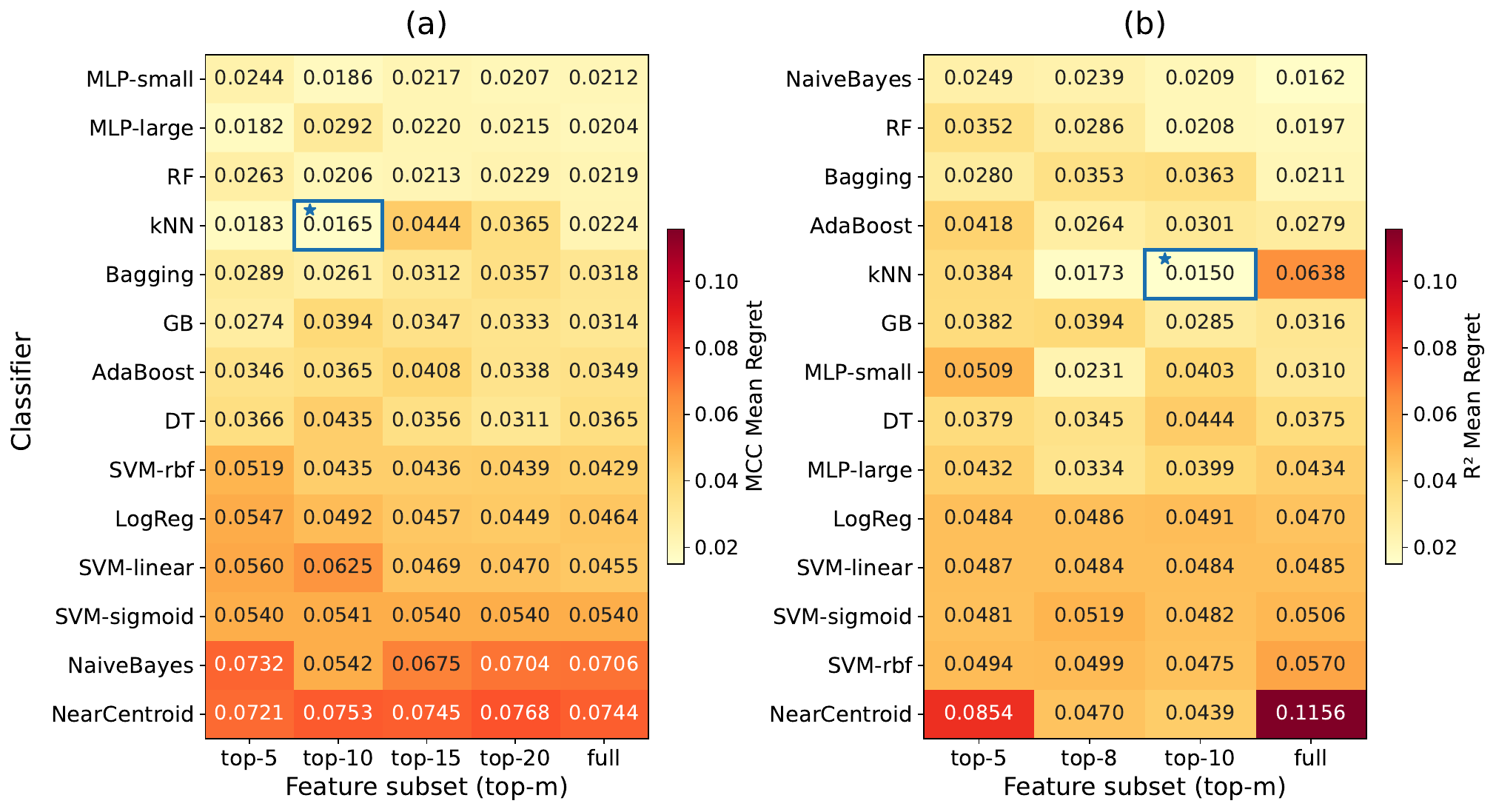}
  \caption{LOO mean regret across all classifier-feature subset configurations for (a)~classification ($\bar\rho^{\rm (MCC)}$) and (b)~regression ($\bar\rho^{(R^2)}$). Lower values (lighter cells) are better. In each panel, the minimum-regret configuration is marked with~$\star$, and classifiers are ordered by ascending overall mean regret within that panel.
  }
  \label{fig:loo_heatmap}
\end{figure*}

\texttt{ModelSelection} searches exhaustively over 14 classical machine learning models (see the y-axis labels in Fig.~\ref{fig:loo_heatmap}). Each configuration is evaluated by leave-one-out cross-validation~(LOO) \cite{1643031.1643047} over the meta-dataset, using mean regret as the selection criterion. We quantify recommendation quality using \emph{regret} \cite{Savage1951}, the performance gap between the recommended circuit and the best-performing circuit for a given dataset, such that zero regret corresponds to an optimal recommendation. Each regret value is a difference computed within its own metric's scale, so zero always means the recommended circuit matches the best achievable one on that scale. In each LOO fold, the recommender is trained on the remaining $N-1$ datasets and evaluated on the held-out dataset $i$, producing a recommendation $\hat{k}_i$ without using dataset $i$ during training. Mean regret is then defined as
\begin{equation}
  \bar{\rho} = \frac{1}{N}\sum_{i=1}^{N}
  \bigl(\mathcal{Y}_{i,k^*_i} - \mathcal{Y}_{i,\hat{k}_i}\bigr),
  \label{eq:regret}
\end{equation}
where $\mathcal{Y}_{i,k^*_i} = \max\limits_{k} \mathcal{Y}_{i,k}$ is the best achievable score on dataset $i$, and $\hat{k}_i$ indexes the top-ranked circuit recommended by the model. Instantiating $\mathcal{Y}$ with $\mathcal{Y}^{(MCC)}$ or $\mathcal{Y}^{(R^2)}$ yields the classification and regression mean regrets $\bar\rho^{(MCC)}$ and $\bar\rho^{(R^2)}$, shown respectively in Fig.~\ref{fig:loo_heatmap}(a) and (b). The configuration with the lowest mean regret is selected as the task default.     Fig.~\ref{fig:loo_heatmap} reports this LOO evaluation for every classifier--feature-subset combination, computed over the full benchmark meta-dataset introduced in Appendix~\ref{appendix:datasets} ($N=105$ classification, $N=86$ regression datasets). These are the model-selection results underlying the default pre-trained recommenders.

For both tasks, $k$NN with top-10 MI-selected features achieves the lowest mean regret - $\bar\rho^{(MCC)} = 0.0165$ for classification, $\bar\rho^{(R^2)} = 0.0150$ for regression, each marked by $\star$ in Fig.~\ref{fig:loo_heatmap}, and is selected as the default configuration for both.

Each selected configuration is refitted on the full meta-dataset and serialized as a \texttt{PairwiseRecommender} object saved in \texttt{importlib.resources}. Users load the default object with a single call and obtain a fitted recommender without training.

\section{\label{sec:usage}Usage Example}

This section demonstrates the main workflow of Qmes through practical examples. We begin with the basic procedure for obtaining circuit recommendations for a new dataset, followed by examples illustrating additional functionality and customization.

\subsection{Basic usage}
The simplest use of Qmes is to obtain a ranked list of encoding circuits for a new dataset using a pre-trained recommender. Listing~\ref{lst:basic} illustrates this workflow for a classification task using the breast cancer dataset from \texttt{scikit-learn}.

\begin{lstlisting}[
    language=Python,
    caption={An example of obtaining a top-3 circuit recommendation for an unseen dataset.},
    label={lst:basic}
]
from sklearn.datasets import load_breast_cancer
from Qmes import get_extractor, load_default_recommender, recommend
X, y = load_breast_cancer(return_X_y=True)
result = recommend(
    X, y,
    extractor=get_extractor("classification"),
    recommender=load_default_recommender("classification"),
    top_k=3,
)
print(result["top_k"])
# ['unit', 'RY', 'HERx']
\end{lstlisting}

The \texttt{recommend} function takes the dataset (\texttt{X, y}), an extractor, and a pre-trained recommender as input. The \texttt{get\_extractor} function constructs a stateless extractor object, while \texttt{load\_default\_recommender} loads a pre-trained \texttt{PairwiseRecommender}. The \texttt{recommend} function preprocesses (\texttt{X, y}), uses the extractor to compress the data into a complexity vector, verifies that the extractor and recommender are consistent in task type and feature names, and forwards the resulting vector to \texttt{recommender.predict()} function. The recommended circuits can optionally be evaluated by the evaluator, as shown in Listing~\ref{lst:evaluator} below.

\begin{lstlisting}[
    language=Python,
    caption={Optional verification of a Recommender against Evaluator scores.},
    label={lst:evaluator}
]
from Qmes import get_evaluator
evaluator = get_evaluator("classification")
scores = {
    c: evaluator.evaluate_circuit(X, y, c)["mean_mcc"]
    for c in result["top_k"]
}
print(scores)
# {'unit': 0.9110729474590961, 'RY': 0.9105354150706099, 'HERx': 0.90636880439311}
\end{lstlisting}

\subsection{\label{sec:advanced}Advanced usage}

Beyond the pre-trained workflow, Qmes allows users to extend the circuit pool and retrain the recommender for customized settings. Listing~\ref{lst:advanced} illustrates this procedure by registering a new encoding circuit and rebuilding the meta-dataset for the enlarged circuit pool.

\begin{lstlisting}[
    language=Python,
    caption={Registering a new circuit and refitting a recommender on the enlarged pool.},
    label={lst:advanced}
]
import Qmes.circuits.registry as registry
from Qmes import get_extractor, get_evaluator, get_recommender
from sklearn.neighbors import KNeighborsClassifier

registry.CIRCUIT_POOL["myCirc"] = my_encoding_fn    # (1) register circuit

extractor = get_extractor("classification")
evaluator = get_evaluator("classification")
meta  = extractor.extract_batch(datasets)           # (2) meta-features, stage (a)
pivot = evaluator.build_pivot(datasets)             # (3) quantum scores, stage (a)

recommender = get_recommender("classification", KNeighborsClassifier(),
                      feature_names=extractor._feature_names)
recommender.fit(meta.loc[pivot.columns].values, pivot)      # (4) refit, stage (b)
recommender.save("my_bundle")                               # (5) reusable via load()
\end{lstlisting}

In Listing~\ref{lst:advanced}, \texttt{datasets} maps dataset names to $(\texttt{X, y})$ pairs. After the new circuit is registered in step~(1), steps (2)-(4) reproduce the two-stage pipeline of Fig.~\ref{fig:workflow} with the enlarged pool: the \texttt{Extractor} computes the meta-features, the \texttt{Evaluator} generates the circuit-performance scores, and the \texttt{Recommender} is refitted on the resulting meta-dataset. The trained recommender is refitted on a new meta-dataset. The result is then saved in step~(5) for subsequent reuse. An Extractor or Evaluator can be loaded from Qmes or defined as shown in Listing~\ref{lst:extractor} and Listing~\ref{lst:evaluator}, while the circuit pool and the underlying classifier are customized directly, as already shown in steps~(1) and~(4) above, without subclassing.
%

\section{\label{sec:validation}Validation}

\subsection{Testing}

The package includes 81 tests, runs in continuous integration across Python 3.10-3.12, and achieves 94\% combined line-and-branch coverage of the core package (\texttt{pytest-cov}; the Qsun simulator and one-off data-generation scripts are excluded). Beyond data shape and type checks, the test suite also validates functional behavior. For example, the \texttt{Recommender} is tested to recover a known assignment pattern when meta-features are permuted in a controlled way. Coverage is further complemented by mutation testing (\texttt{mutmut}) over the three modules: 631 of 700 mutants are killed, corresponding to a mutation score of 90.1\%, with no survivor in the leave-one-out selection module. The remaining 69 survivors were inspected manually and confirmed to be equivalent mutants.

\subsection{\label{validity} Scientific Validity}

\begin{figure}[t]
  \centering
  \includegraphics[width=\linewidth]{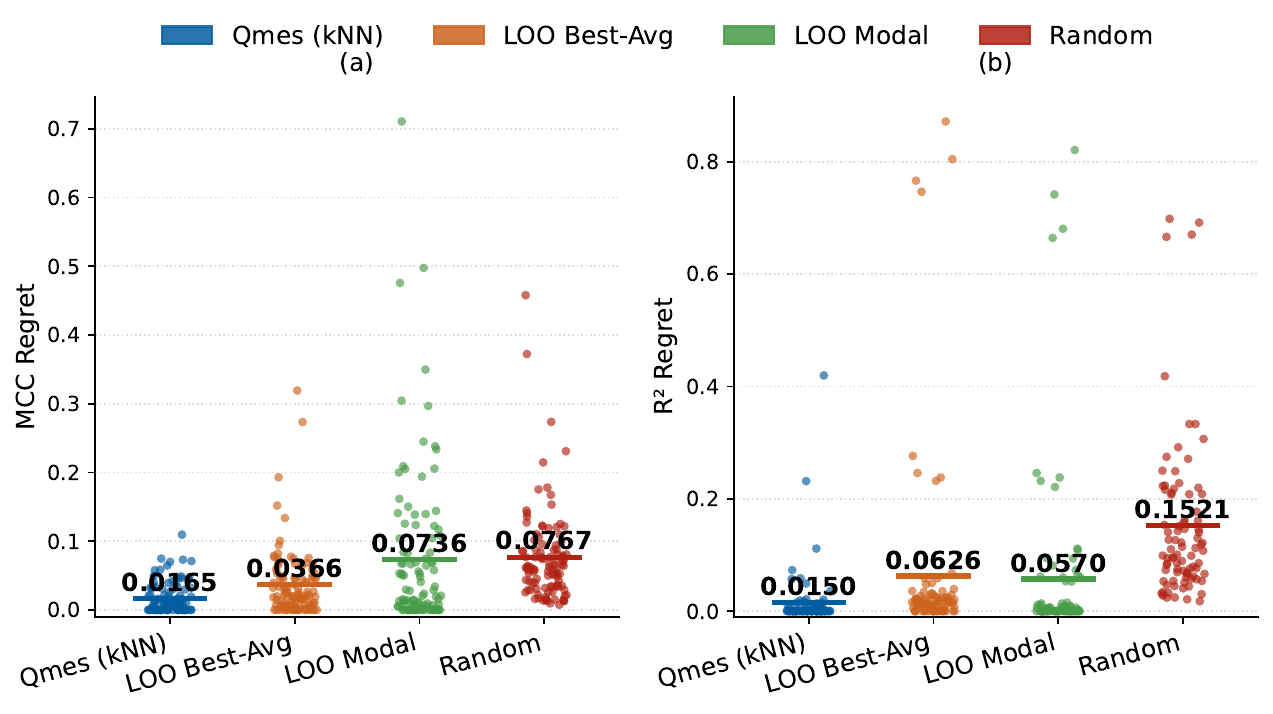}
  \caption{Per-dataset regret $\rho_i$ of the Qmes $k$NN Recommender compared with the three baselines above, for (a) classification (MCC regret) and (b) regression ($R^2$ regret). Each point is one dataset; horizontal bars mark the mean regret $\bar\rho$ per group (annotated).}
  \label{fig:validity_strip}
\end{figure}

To assess the effectiveness of Qmes's recommendations, we compare its performance against three strategies: LOO Best-Avg, LOO Modal, and Random. Table~\ref{tab:baseline_example} first illustrates these strategies on a small hypothetical example; Fig.~\ref{fig:validity_strip} then applies them to the actual benchmark meta-dataset ($N=105$ classification, $N=86$ regression datasets) to produce the real per-dataset comparison against Qmes.
\begin{table}[t]
  \caption{
  Example performance scores for a small hypothetical pool of three circuits ($c_1,c_2,c_3$) over four hypothetical datasets ($D_1$--$D_4$), with invented scores chosen so that the best circuit differs across datasets ($D_2$ favors $c_2$; the rest favor $c_1$). The best-performing circuit for each dataset is shown in bold.}
  \label{tab:baseline_example}
  \begin{tabular*}{\linewidth}{@{\extracolsep{\fill}}lccccc@{}}
    \toprule
    \textbf{Circuit} & $D_1$ & $D_2$ & $D_3$ & $D_4$ & \textbf{mean} \\
    \midrule
    $c_1$ & \textbf{0.90} & 0.80          & \textbf{0.85} & \textbf{0.70} & \textbf{0.81} \\
    $c_2$ & 0.50          & \textbf{0.90} & 0.40          & 0.50          & 0.58 \\
    $c_3$ & 0.40          & 0.30          & 0.60          & 0.60          & 0.48 \\
    \bottomrule
  \end{tabular*}
\end{table}

\emph{LOO Best-Avg} recommends, for each held-out dataset, the circuit with the highest mean score over the remaining datasets. For example, holding out $D_2$, $c_1$'s mean over $D_1, D_3, D_4$ (0.82) exceeds $c_2$'s (0.47) and $c_3$'s (0.53), so $c_1$ is recommended even though $c_2$ is actually best on $D_2$, incurring a regret of $0.10$. On $D_1$, $D_3$, and $D_4$, $c_1$ is both the \emph{LOO Best-Avg} pick and the true best circuit, so regret is zero there. \emph{LOO Modal} recommends, for each held-out dataset, whichever circuit wins on the most of the remaining datasets; here, that is $c_1$ in every fold, so \emph{LOO Modal} incurs the same regrets as \emph{LOO Best-Avg}. \emph{Random} just picks a circuit uniformly at random from the pool, so its regret on a dataset is the average regret across all circuits in the pool, e.g., on $D_1$, averaging each circuit's gap from the best (0, 0.40, 0.50) gives a regret of $0.30$.

As Fig.~\ref{fig:validity_strip} shows, Qmes reduces mean regret from $\bar\rho=0.0366$ (LOO Best-Avg) to $0.0165$ for classification and from $0.0626$ to $0.0150$ for regression, a $2.2\times$ and $4.2\times$ reduction, respectively. To verify that these reductions are systematic rather than driven by a few favorable datasets, we apply a paired Wilcoxon signed-rank test~\cite{JMLR:v7:demsar06a} to the per-dataset regret differences between Qmes and the LOO Best-Avg baseline. The test compares two methods evaluated on the same collection of datasets, asking whether one is consistently better across them rather than better on average; the improvement is significant for both tasks ($p = 5.42\times10^{-5}$ for classification, $p = 1.56\times10^{-5}$ for regression).

\subsection{\label{sec:htru2} External Case Study: HTRU2}

We further apply Qmes's default classification recommender to HTRU2~\cite{htru2_372}, a pulsar candidate dataset from the High Time Resolution Universe survey. A pulsar is a rapidly rotating neutron star whose beamed radio emission reaches the observer as a periodic signal. Each detection returned by a survey pipeline is a \emph{candidate} that must be labelled either a genuine pulsar or radio frequency interference/noise, which makes candidate selection a binary classification problem. HTRU2 contains $17{,}898$ human-annotated candidates, of which $1{,}639$ are real pulsars and $16{,}259$ are spurious, and describes each candidate by eight continuous features: four summary statistics (mean, standard deviation, excess kurtosis and skewness) of the integrated pulse profile, and the same four statistics of the DM--SNR curve.

On a stratified subsample of $600$ candidates ($55$ pulsars), the recommender returns \texttt{HERx}, \texttt{RY} and \texttt{unit}  from HTRU2's complexity vector. Across all three encodings the FP column of Table~\ref{tab:htru2} is zero, all $545$ non-pulsars are classified correctly, so the circuits differ only in recall: how many of the $55$ true pulsars each one recovers. \texttt{HERx} is the strongest, reaching $\mathcal{Y}^{(\text{MCC})}=0.862$: it recovers $42$ of the $55$ pulsars and misses $13$. \texttt{RY} follows at $0.829$ ($39$ recovered, $16$ missed), and \texttt{unit} at $0.781$ ($35$ recovered, $20$ missed).
\begin{table}[t]
  \centering
  \caption{Circuits recommended by Qmes for HTRU2 and their held-out confusion
  matrices, with pulsar as the positive class.}
  \label{tab:htru2}
  \begin{tabular*}{\linewidth}{@{\extracolsep{\fill}}lccccc@{}}
    \toprule
    \textbf{Qmes circuit} & $\mathcal{Y}^{(\text{MCC})}$ & \textbf{TN} & \textbf{FP} & \textbf{FN} & \textbf{TP} \\
    \midrule
    \texttt{HERx} & 0.862 & 545 & 0 & 13 & 42 \\
    \texttt{RY}   & 0.829 & 545 & 0 & 16 & 39 \\
    \texttt{unit} & 0.781 & 545 & 0 & 20 & 35 \\
    \bottomrule
  \end{tabular*}
\end{table}

\begin{figure}[b]
  \centering
  \includegraphics[width=\linewidth]{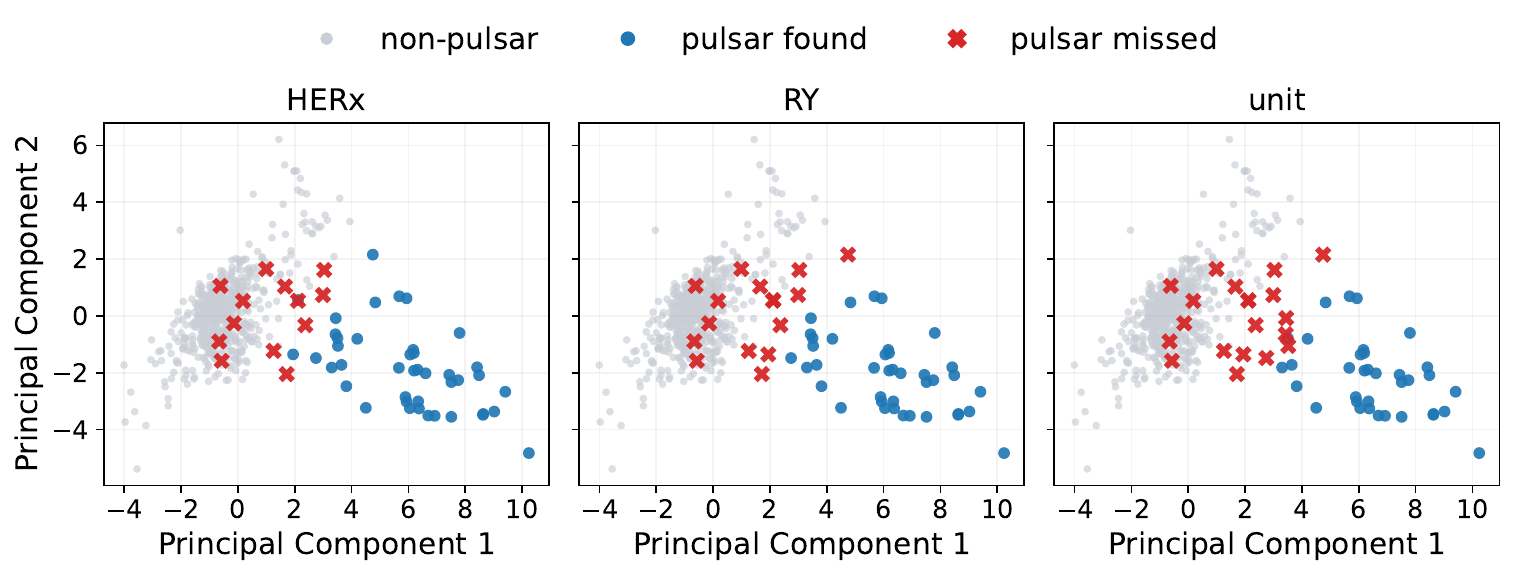}
  \caption{Classification of the HTRU2 candidates under each recommended encoding. One panel per circuit; each point is one candidate: grey marks a non-pulsar, blue a pulsar that is classified correctly, and a red cross a pulsar that is misclassified as noise. The axes are the first two principal components of the standardised features.}
  \label{fig:htru2}
\end{figure}
Figure~\ref{fig:htru2} explains the difference by showing where each circuit's misclassified pulsars fall. The non-pulsars (grey) form one dense cluster; the
true pulsars extend from its edge toward the lower right. Pulsars in the part of the plane with no non-pulsars are well separated, and every circuit classifies
them correctly (blue in all panels). The misclassified pulsars (red crosses, the FN entries of Table~\ref{tab:htru2}) lie instead in the class-overlap region, among the non-pulsars and close to the decision boundary. Because every circuit recovers the well-separated pulsars, the $\mathcal{Y}^{(\text{MCC})}$ ranking is decided entirely in this overlap region.
\section{\label{sec:conclusion}Conclusion}

Qmes turns circuit selection from a per-dataset quantum evaluation loop into a single classical inference call: obtaining a recommendation requires only meta-feature extraction and a query to $f$, without additional quantum evaluation and no need to reimplement the offline evaluation pipeline. Beyond the meta-learning framework we previously proposed \cite{tung2026automatedselectionquantumencoding}, Qmes extends the approach to regression pipeline alongside classification, a one-vs-one pairwise reformulation suited to the small-$N$ meta-dataset, and pre-trained recommenders. In leave-one-out validation over $105$ classification and $86$ regression datasets, Qmes roughly halves the mean regret to the best circuit for classification and reduces it to under a quarter for regression against the non-adaptive baseline.

Several limitations bound the current scope. First, Qmes targets quantum kernel methods with fixed, non-trainable circuits. Second, meta-features are computed on at most 600 subsampled data points, so for larger real-world datasets the complexity vector characterizes a subsample rather than the full data. Third, all circuit scores derive from noiseless state-vector simulation, so hardware errors are not reflected in the recommendations. The architecture supports several extensions we intend to pursue. Both the circuit pool and the meta-dataset can be enlarged without structural changes. The evaluator interface admits two extensions: a variational oracle in which circuit parameters are trained rather than fixed, and evaluation under realistic noise, via noisy simulation or real hardware. Finally, extending Qmes beyond supervised learning to unsupervised tasks would require suitable meta-features that do not rely on labelled targets.

\section*{CRediT authorship contribution statement}

D.D.T. initiated the study, developed the theoretical framework and simulation code, conducted the experiments and analyzed the results, and prepared the initial manuscript. Q.C.N. and V.T.H. contributed to the methodology, supported the validation of the numerical findings, and participated in manuscript revision. L.B.H. and L.N.T. provided overall supervision, obtained funding for the project, and contributed to the revision of the manuscript. All authors discussed the findings, reviewed the manuscript, and approved the final version.

\section*{Declaration of competing interest}
The authors declare that they have no known competing financial interests or personal
relationships that could have appeared to influence the work reported in this paper.

\section*{Acknowledgements}

L. B. H is funded by the Tohoku Initiative for Fostering Global Researchers for Interdisciplinary Sciences (TI-FRIS) of MEXT's Strategic Professional Development Program for Young Researchers. V.T.H is partially supported by the VNUHCM - University of Information Technology's Scientific Research Support Fund.

\section*{Data availability}
The code and data are available on GitHub at \url{https://github.com/tungduy1704/Qmes}.


\appendix
\section{\label{appendix:features} Complexity metrics}
The 22 classification meta-features used by \texttt{ClassificationExtractor} are listed by category in Ref.~\cite{tung2026automatedselectionquantumencoding}. Table~\ref{tab:complexity_metrics_reg} lists the 12 regression meta-features used by \texttt{RegressionExtractor} (Sec.~\ref{extractor}).
\begin{table*}[t]
    \centering
    \caption{Regression descriptors.}
    \label{tab:complexity_metrics_reg}
    \footnotesize
    \renewcommand{\arraystretch}{1.3}
    \begin{tabular}{@{}lll@{}}
        \toprule
        \textbf{Category} & \textbf{Metrics} & \textbf{Description} \\
        \midrule
        \parbox{2.8cm}{\raggedright Correlation}
        & \parbox{3cm}{\raggedright $c_1, c_2, c_3$ \cite{990132, Lorena2018}, $c_4$\cite{Lorena2018}}
        & \parbox{10cm}{\raggedright Assess how strongly individual features relate to the output, and how much of the data can be explained through such relationships, via rank correlation and correlation-guided example elimination.} \\
        \midrule
        \parbox{2.8cm}{\raggedright Linearity}
        & \parbox{3cm}{\raggedright $l_1, l_2$ \cite{990132, Lorena2018}}
        & \parbox{10cm}{\raggedright Measure how well a linear function fits the data, via the residual error of a multivariate linear regression.} \\
        \midrule
        \parbox{2.8cm}{\raggedright Smoothness}
        & \parbox{3cm}{\raggedright $s_1, s_2, s_3$ \cite{990132, Lorena2018}}
        & \parbox{10cm}{\raggedright Assess whether nearby points in the input space also have similar output values, using minimum-spanning-tree distances and nearest-neighbor prediction error.} \\
        \midrule
        \parbox{2.8cm}{\raggedright Geometry}
        & \parbox{3cm}{\raggedright $l_3$ \cite{990132}, $s_4$ \cite{Lorena2018}, $t_2$ \cite{990132}}
        & \parbox{10cm}{\raggedright Capture the spatial structure of the data by measuring model sensitivity to synthetically interpolated points and the ratio of samples to feature dimensionality.} \\
        \bottomrule
    \end{tabular}
\end{table*}

\section{\label{appendix:datasets}Dataset list}

The meta-dataset comprises $N=105$ classification and $N=86$ regression benchmark datasets, drawn from two categories: \emph{real-world} datasets (sklearn built-ins, the UCI ML Repository, and bundled CSV files) and \emph{synthetic} datasets generated from sklearn and custom generators. The complete registry, including UCI repository IDs and per-dataset generator parameters sufficient for exact reproduction, is given in \texttt{Qmes/data/clf/train.py} and \texttt{Qmes/data/reg/train.py}.

\subsection{Real-world datasets}

Table~\ref{tab:datasets_real} summarizes the 48 real-world classification and 23 real-world regression datasets by source. UCI datasets span diverse domains (medical diagnosis, biometric authentication, financial credit scoring, materials science) and are drawn from the UCI ML Repository via \texttt{ucimlrepo.fetch\_ucirepo}.

\begin{table*}[t]
  \centering
  \caption{Real-world datasets by source. Representative examples are
  illustrative, not exhaustive; sklearn multiclass datasets (Iris, Wine,
  Digits) are each split into multiple binary-class pairs, counted
  separately in the registry.}
  \label{tab:datasets_real}
  \footnotesize
  \begin{tabular}{@{}llcl@{}}
    \toprule
    \textbf{Task} & \textbf{Source} & \textbf{\#Datasets} & \textbf{Representative examples} \\
    \midrule
    Classification & sklearn & 12 & Iris, Wine, Digits, Breast Cancer \\
    Classification & UCI     & 34 & Ionosphere, Mushroom, Wdbc, Adult Income, Sonar \\
    Classification & CSV     & 2  & BankNote Auth, Pima Diabetes \\
    \midrule
    Regression     & sklearn & 2  & Diabetes, California Housing \\
    Regression     & UCI     & 21 & Abalone, Concrete, Energy Efficiency, Bike Sharing, Wine Quality Red \\
    \bottomrule
  \end{tabular}
\end{table*}

\subsection{Synthetic datasets}

The remaining 57 classification and 63 regression datasets are generated programmatically as parameter configurations of a small number of generator families, summarized in Table~\ref{tab:datasets_synth}. For classification, four generators are sklearn built-ins (\texttt{make\_blobs}, \texttt{make\_moons}, \texttt{make\_circles}, \texttt{make\_classification}) and five are custom (XOR, checkerboard, two-spirals, nonlinear-subspace, rotated-moons), covering varied noise, class imbalance, feature redundancy, and embedding
dimensionality. For regression, four generators are sklearn built-ins (\texttt{make\_regression} and the three Friedman benchmarks) and six are custom (quadratic, sinusoidal, interaction, heteroscedastic, outlier-contaminated, correlated-input).

\begin{table*}[htbp]
  \centering
  \caption{Synthetic classification datasets by generator family. (Top) $n_{\text{features}}$ fixed where the generator exposes no
  feature-count parameter. (Bottom) $n_{\text{features}}$ fixed at 4 for Friedman-2/3
  (sklearn-defined)}
  \label{tab:datasets_synth}
  \footnotesize
  \begin{tabular*}{\textwidth}{@{\extracolsep{\fill}}llccl@{}}
    \toprule
    \textbf{Generator} & \textbf{Source} & \textbf{\#Datasets} & \textbf{$n_{\text{samples}}$} & \textbf{$n_{\text{features}}$} \\
    \midrule
    \texttt{blobs}           & sklearn & 5  & 200--300 & 2--8 \\
    \texttt{classification}  & sklearn & 25 & 200--1000 & 4--30 \\
    \texttt{moons}           & sklearn & 4  & 200--500 & 2 (fixed) \\
    \texttt{circles}         & sklearn & 3  & 200--300 & 2 (fixed) \\
    \texttt{xor}             & custom  & 5  & 300      & 2--6 \\
    \texttt{checkerboard}    & custom  & 4  & 300--400 & 2 (fixed) \\
    \texttt{two\_spirals}    & custom  & 4  & 300--500 & 2 (fixed) \\
    \texttt{nonlin\_subspace}& custom  & 3  & 300      & 4--8 \\
    \texttt{moons\_rotated}  & custom  & 4  & 300      & 3--7 \\ \midrule\midrule
    \texttt{regression}      & sklearn & 22 & 30--400  & 2--30 \\
    \texttt{friedman1}       & sklearn & 4  & 300      & 5--15 \\
    \texttt{friedman2}       & sklearn & 2  & 300      & 4 (fixed) \\
    \texttt{friedman3}       & sklearn & 2  & 300      & 4 (fixed) \\
    quadratic                & custom  & 8  & 300--400 & 2--12 \\
    sinusoidal               & custom  & 5  & 300      & 3--4 \\
    interaction              & custom  & 8  & 300--400 & 4--10 \\
    heteroscedastic          & custom  & 5  & 100--300 & 4--10 \\
    outlier                  & custom  & 1  & 300      & 4 \\
    correlated               & custom  & 6  & 300      & 5--10 \\
    \bottomrule
  \end{tabular*}
\end{table*}

\section{\label{circuit} Circuit Pool}

Qmes includes a fixed pool of circuits (Table~\ref{tab:circuits}), spanning amplitude embeddings, separable single-qubit rotations, and entangling circuits, all implemented in the bundled Qsun simulator and exposed through \texttt{CIRCUIT\_POOL} and \texttt{get\_circuit\_names()}. This pool can be extended easily in future versions.

\begin{table*}[htbp]
  \caption{The seven circuits in the Qmes pool.}
  \label{tab:circuits}
  \footnotesize
    \begin{tabular*}{\textwidth}{@{\extracolsep{\fill}}llll@{}}
        \toprule
        \textbf{Name} & \textbf{Encoding} & \textbf{2-qubit gate} & \textbf{Features/Qubits} \\
        \midrule
        \texttt{unit}   & Square-root amplitude (per-qubit) & None         & 4  \\
        \texttt{SRx}    & Separable RX                      & None         & 4  \\
        \texttt{RY}     & Angle Encoding (RY)                & None         & 4  \\
        \texttt{HERx}   & Hardware-Efficient RX               & Linear (CX)  & 4  \\
        \texttt{RY\_CX} & Angle (RY) + Linear CX              & Linear (CX)  & 4  \\
        \texttt{ZFM}    & Z Feature Map                       & None         & 4  \\
        \texttt{HD}     & High-Dimensional (RZ-RY-RZ)          & Brickwork (ISWAP) & 4 \\
        \bottomrule
    \end{tabular*}
\end{table*}

\bibliography{citation}

\begin{thebibliography}{29}%
\makeatletter
\providecommand \@ifxundefined [1]{%
 \@ifx{#1\undefined}
}%
\providecommand \@ifnum [1]{%
 \ifnum #1\expandafter \@firstoftwo
 \else \expandafter \@secondoftwo
 \fi
}%
\providecommand \@ifx [1]{%
 \ifx #1\expandafter \@firstoftwo
 \else \expandafter \@secondoftwo
 \fi
}%
\providecommand \natexlab [1]{#1}%
\providecommand \enquote  [1]{``#1''}%
\providecommand \bibnamefont  [1]{#1}%
\providecommand \bibfnamefont [1]{#1}%
\providecommand \citenamefont [1]{#1}%
\providecommand \href@noop [0]{\@secondoftwo}%
\providecommand \href [0]{\begingroup \@sanitize@url \@href}%
\providecommand \@href[1]{\@@startlink{#1}\@@href}%
\providecommand \@@href[1]{\endgroup#1\@@endlink}%
\providecommand \@sanitize@url [0]{\catcode `\\12\catcode `\$12\catcode
  `\&12\catcode `\#12\catcode `\^12\catcode `\_12\catcode `\%12\relax}%
\providecommand \@@startlink[1]{}%
\providecommand \@@endlink[0]{}%
\providecommand \url  [0]{\begingroup\@sanitize@url \@url }%
\providecommand \@url [1]{\endgroup\@href {#1}{\urlprefix }}%
\providecommand \urlprefix  [0]{URL }%
\providecommand \Eprint [0]{\href }%
\providecommand \doibase [0]{https://doi.org/}%
\providecommand \selectlanguage [0]{\@gobble}%
\providecommand \bibinfo  [0]{\@secondoftwo}%
\providecommand \bibfield  [0]{\@secondoftwo}%
\providecommand \translation [1]{[#1]}%
\providecommand \BibitemOpen [0]{}%
\providecommand \bibitemStop [0]{}%
\providecommand \bibitemNoStop [0]{.\EOS\space}%
\providecommand \EOS [0]{\spacefactor3000\relax}%
\providecommand \BibitemShut  [1]{\csname bibitem#1\endcsname}%
\let\auto@bib@innerbib\@empty
\bibitem [{\citenamefont {Lloyd}\ \emph {et~al.}(2014)\citenamefont {Lloyd},
  \citenamefont {Mohseni},\ and\ \citenamefont {Rebentrost}}]{Lloyd2014}%
  \BibitemOpen
  \bibfield  {author} {\bibinfo {author} {\bibfnamefont {S.}~\bibnamefont
  {Lloyd}}, \bibinfo {author} {\bibfnamefont {M.}~\bibnamefont {Mohseni}},\
  and\ \bibinfo {author} {\bibfnamefont {P.}~\bibnamefont {Rebentrost}},\
  }\bibfield  {title} {\bibinfo {title} {Quantum principal component
  analysis},\ }\href {https://doi.org/10.1038/nphys3029} {\bibfield  {journal}
  {\bibinfo  {journal} {Nature Physics}\ }\textbf {\bibinfo {volume} {10}},\
  \bibinfo {pages} {631} (\bibinfo {year} {2014})}\BibitemShut {NoStop}%
\bibitem [{\citenamefont {Rebentrost}\ \emph {et~al.}(2014)\citenamefont
  {Rebentrost}, \citenamefont {Mohseni},\ and\ \citenamefont
  {Lloyd}}]{PhysRevLett.113.130503}%
  \BibitemOpen
  \bibfield  {author} {\bibinfo {author} {\bibfnamefont {P.}~\bibnamefont
  {Rebentrost}}, \bibinfo {author} {\bibfnamefont {M.}~\bibnamefont
  {Mohseni}},\ and\ \bibinfo {author} {\bibfnamefont {S.}~\bibnamefont
  {Lloyd}},\ }\bibfield  {title} {\bibinfo {title} {Quantum support vector
  machine for big data classification},\ }\href
  {https://doi.org/10.1103/PhysRevLett.113.130503} {\bibfield  {journal}
  {\bibinfo  {journal} {Phys. Rev. Lett.}\ }\textbf {\bibinfo {volume} {113}},\
  \bibinfo {pages} {130503} (\bibinfo {year} {2014})}\BibitemShut {NoStop}%
\bibitem [{\citenamefont {Lloyd}\ \emph {et~al.}(2016)\citenamefont {Lloyd},
  \citenamefont {Garnerone},\ and\ \citenamefont {Zanardi}}]{Lloyd2016}%
  \BibitemOpen
  \bibfield  {author} {\bibinfo {author} {\bibfnamefont {S.}~\bibnamefont
  {Lloyd}}, \bibinfo {author} {\bibfnamefont {S.}~\bibnamefont {Garnerone}},\
  and\ \bibinfo {author} {\bibfnamefont {P.}~\bibnamefont {Zanardi}},\
  }\bibfield  {title} {\bibinfo {title} {Quantum algorithms for topological and
  geometric analysis of data},\ }\href {https://doi.org/10.1038/ncomms10138}
  {\bibfield  {journal} {\bibinfo  {journal} {Nature Communications}\ }\textbf
  {\bibinfo {volume} {7}},\ \bibinfo {pages} {10138} (\bibinfo {year}
  {2016})}\BibitemShut {NoStop}%
\bibitem [{\citenamefont {Lloyd}\ \emph {et~al.}(2020)\citenamefont {Lloyd},
  \citenamefont {Schuld}, \citenamefont {Ijaz}, \citenamefont {Izaac},\ and\
  \citenamefont {Killoran}}]{lloyd2020quantumembeddingsmachinelearning}%
  \BibitemOpen
  \bibfield  {author} {\bibinfo {author} {\bibfnamefont {S.}~\bibnamefont
  {Lloyd}}, \bibinfo {author} {\bibfnamefont {M.}~\bibnamefont {Schuld}},
  \bibinfo {author} {\bibfnamefont {A.}~\bibnamefont {Ijaz}}, \bibinfo {author}
  {\bibfnamefont {J.}~\bibnamefont {Izaac}},\ and\ \bibinfo {author}
  {\bibfnamefont {N.}~\bibnamefont {Killoran}},\ }\href
  {https://arxiv.org/abs/2001.03622} {\bibinfo {title} {Quantum embeddings for
  machine learning}} (\bibinfo {year} {2020}),\ \Eprint
  {https://arxiv.org/abs/2001.03622} {arXiv:2001.03622 [quant-ph]} \BibitemShut
  {NoStop}%
\bibitem [{\citenamefont {Havl{\'i}{\v{c}}ek}\ \emph
  {et~al.}(2019)\citenamefont {Havl{\'i}{\v{c}}ek}, \citenamefont
  {C{\'o}rcoles}, \citenamefont {Temme}, \citenamefont {Harrow}, \citenamefont
  {Kandala}, \citenamefont {Chow},\ and\ \citenamefont
  {Gambetta}}]{Havlicek2019}%
  \BibitemOpen
  \bibfield  {author} {\bibinfo {author} {\bibfnamefont {V.}~\bibnamefont
  {Havl{\'i}{\v{c}}ek}}, \bibinfo {author} {\bibfnamefont {A.~D.}\ \bibnamefont
  {C{\'o}rcoles}}, \bibinfo {author} {\bibfnamefont {K.}~\bibnamefont {Temme}},
  \bibinfo {author} {\bibfnamefont {A.~W.}\ \bibnamefont {Harrow}}, \bibinfo
  {author} {\bibfnamefont {A.}~\bibnamefont {Kandala}}, \bibinfo {author}
  {\bibfnamefont {J.~M.}\ \bibnamefont {Chow}},\ and\ \bibinfo {author}
  {\bibfnamefont {J.~M.}\ \bibnamefont {Gambetta}},\ }\bibfield  {title}
  {\bibinfo {title} {Supervised learning with quantum-enhanced feature
  spaces},\ }\href {https://doi.org/10.1038/s41586-019-0980-2} {\bibfield
  {journal} {\bibinfo  {journal} {Nature}\ }\textbf {\bibinfo {volume} {567}},\
  \bibinfo {pages} {209} (\bibinfo {year} {2019})}\BibitemShut {NoStop}%
\bibitem [{\citenamefont {Peters}\ \emph {et~al.}(2021)\citenamefont {Peters},
  \citenamefont {Caldeira}, \citenamefont {Ho}, \citenamefont {Leichenauer},
  \citenamefont {Mohseni}, \citenamefont {Neven}, \citenamefont {Spentzouris},
  \citenamefont {Strain},\ and\ \citenamefont {Perdue}}]{Peters2021}%
  \BibitemOpen
  \bibfield  {author} {\bibinfo {author} {\bibfnamefont {E.}~\bibnamefont
  {Peters}}, \bibinfo {author} {\bibfnamefont {J.}~\bibnamefont {Caldeira}},
  \bibinfo {author} {\bibfnamefont {A.}~\bibnamefont {Ho}}, \bibinfo {author}
  {\bibfnamefont {S.}~\bibnamefont {Leichenauer}}, \bibinfo {author}
  {\bibfnamefont {M.}~\bibnamefont {Mohseni}}, \bibinfo {author} {\bibfnamefont
  {H.}~\bibnamefont {Neven}}, \bibinfo {author} {\bibfnamefont
  {P.}~\bibnamefont {Spentzouris}}, \bibinfo {author} {\bibfnamefont
  {D.}~\bibnamefont {Strain}},\ and\ \bibinfo {author} {\bibfnamefont {G.~N.}\
  \bibnamefont {Perdue}},\ }\bibfield  {title} {\bibinfo {title} {Machine
  learning of high dimensional data on a noisy quantum processor},\ }\href
  {https://doi.org/10.1038/s41534-021-00498-9} {\bibfield  {journal} {\bibinfo
  {journal} {npj Quantum Information}\ }\textbf {\bibinfo {volume} {7}},\
  \bibinfo {pages} {161} (\bibinfo {year} {2021})}\BibitemShut {NoStop}%
\bibitem [{\citenamefont
  {Schuld}(2021)}]{schuld2021supervisedquantummachinelearning}%
  \BibitemOpen
  \bibfield  {author} {\bibinfo {author} {\bibfnamefont {M.}~\bibnamefont
  {Schuld}},\ }\href {https://arxiv.org/abs/2101.11020} {\bibinfo {title}
  {Supervised quantum machine learning models are kernel methods}} (\bibinfo
  {year} {2021}),\ \Eprint {https://arxiv.org/abs/2101.11020} {arXiv:2101.11020
  [quant-ph]} \BibitemShut {NoStop}%
\bibitem [{\citenamefont {Mitarai}\ \emph {et~al.}(2018)\citenamefont
  {Mitarai}, \citenamefont {Negoro}, \citenamefont {Kitagawa},\ and\
  \citenamefont {Fujii}}]{PhysRevA.98.032309}%
  \BibitemOpen
  \bibfield  {author} {\bibinfo {author} {\bibfnamefont {K.}~\bibnamefont
  {Mitarai}}, \bibinfo {author} {\bibfnamefont {M.}~\bibnamefont {Negoro}},
  \bibinfo {author} {\bibfnamefont {M.}~\bibnamefont {Kitagawa}},\ and\
  \bibinfo {author} {\bibfnamefont {K.}~\bibnamefont {Fujii}},\ }\bibfield
  {title} {\bibinfo {title} {Quantum circuit learning},\ }\href
  {https://doi.org/10.1103/PhysRevA.98.032309} {\bibfield  {journal} {\bibinfo
  {journal} {Phys. Rev. A}\ }\textbf {\bibinfo {volume} {98}},\ \bibinfo
  {pages} {032309} (\bibinfo {year} {2018})}\BibitemShut {NoStop}%
\bibitem [{\citenamefont {Schuld}\ and\ \citenamefont
  {Killoran}(2019)}]{PhysRevLett.122.040504}%
  \BibitemOpen
  \bibfield  {author} {\bibinfo {author} {\bibfnamefont {M.}~\bibnamefont
  {Schuld}}\ and\ \bibinfo {author} {\bibfnamefont {N.}~\bibnamefont
  {Killoran}},\ }\bibfield  {title} {\bibinfo {title} {Quantum machine learning
  in feature hilbert spaces},\ }\href
  {https://doi.org/10.1103/PhysRevLett.122.040504} {\bibfield  {journal}
  {\bibinfo  {journal} {Phys. Rev. Lett.}\ }\textbf {\bibinfo {volume} {122}},\
  \bibinfo {pages} {040504} (\bibinfo {year} {2019})}\BibitemShut {NoStop}%
\bibitem [{\citenamefont {Jha}\ \emph {et~al.}(2026)\citenamefont {Jha},
  \citenamefont {Kasabov}, \citenamefont {Bhattacharyya}, \citenamefont
  {Coyle},\ and\ \citenamefont {Prasad}}]{Jha2026}%
  \BibitemOpen
  \bibfield  {author} {\bibinfo {author} {\bibfnamefont {R.~K.}\ \bibnamefont
  {Jha}}, \bibinfo {author} {\bibfnamefont {N.}~\bibnamefont {Kasabov}},
  \bibinfo {author} {\bibfnamefont {S.}~\bibnamefont {Bhattacharyya}}, \bibinfo
  {author} {\bibfnamefont {D.}~\bibnamefont {Coyle}},\ and\ \bibinfo {author}
  {\bibfnamefont {G.}~\bibnamefont {Prasad}},\ }\bibfield  {title} {\bibinfo
  {title} {Comparative performance analysis of quantum feature maps for quantum
  kernel-based machine learning},\ }\href
  {https://doi.org/10.1038/s41598-026-39392-9} {\bibfield  {journal} {\bibinfo
  {journal} {Scientific Reports}\ }\textbf {\bibinfo {volume} {16}},\ \bibinfo
  {pages} {8142} (\bibinfo {year} {2026})}\BibitemShut {NoStop}%
\bibitem [{\citenamefont {Incudini}\ \emph {et~al.}(2024)\citenamefont
  {Incudini}, \citenamefont {Bosco}, \citenamefont {Martini}, \citenamefont
  {Grossi}, \citenamefont {Serra},\ and\ \citenamefont {Pierro}}]{10812182}%
  \BibitemOpen
  \bibfield  {author} {\bibinfo {author} {\bibfnamefont {M.}~\bibnamefont
  {Incudini}}, \bibinfo {author} {\bibfnamefont {D.~L.}\ \bibnamefont {Bosco}},
  \bibinfo {author} {\bibfnamefont {F.}~\bibnamefont {Martini}}, \bibinfo
  {author} {\bibfnamefont {M.}~\bibnamefont {Grossi}}, \bibinfo {author}
  {\bibfnamefont {G.}~\bibnamefont {Serra}},\ and\ \bibinfo {author}
  {\bibfnamefont {A.~D.}\ \bibnamefont {Pierro}},\ }\bibfield  {title}
  {\bibinfo {title} {Automatic and effective discovery of quantum kernels},\
  }\href {https://doi.org/10.1109/TETCI.2024.3499993} {\bibfield  {journal}
  {\bibinfo  {journal} {IEEE Transactions on Emerging Topics in Computational
  Intelligence}\ ,\ \bibinfo {pages} {1}} (\bibinfo {year} {2024})}\BibitemShut
  {NoStop}%
\bibitem [{\citenamefont
  {de~Paula~Neto}(2025)}]{neto2025datacomplexitymeasuresquantum}%
  \BibitemOpen
  \bibfield  {author} {\bibinfo {author} {\bibfnamefont {F.~M.}\ \bibnamefont
  {de~Paula~Neto}},\ }\href
  {https://doi.org/https://doi.org/10.48550/arXiv.2502.15129} {\bibinfo {title}
  {Data complexity measures for quantum circuits architecture recommendation}}
  (\bibinfo {year} {2025})\BibitemShut {NoStop}%
\bibitem [{\citenamefont {Tung}\ \emph {et~al.}(2026)\citenamefont {Tung},
  \citenamefont {Chuong}, \citenamefont {Hai}, \citenamefont {Ho},\ and\
  \citenamefont {Tran}}]{tung2026automatedselectionquantumencoding}%
  \BibitemOpen
  \bibfield  {author} {\bibinfo {author} {\bibfnamefont {D.~D.}\ \bibnamefont
  {Tung}}, \bibinfo {author} {\bibfnamefont {N.~Q.}\ \bibnamefont {Chuong}},
  \bibinfo {author} {\bibfnamefont {V.~T.}\ \bibnamefont {Hai}}, \bibinfo
  {author} {\bibfnamefont {L.~B.}\ \bibnamefont {Ho}},\ and\ \bibinfo {author}
  {\bibfnamefont {L.~N.}\ \bibnamefont {Tran}},\ }\href
  {https://arxiv.org/abs/2604.19076} {\bibinfo {title} {Towards automated
  selection of quantum encoding circuits via meta-learning}} (\bibinfo {year}
  {2026}),\ \Eprint {https://arxiv.org/abs/2604.19076} {arXiv:2604.19076
  [quant-ph]} \BibitemShut {NoStop}%
\bibitem [{\citenamefont {Huang}\ \emph {et~al.}(2021)\citenamefont {Huang},
  \citenamefont {Broughton}, \citenamefont {Mohseni}, \citenamefont {Babbush},
  \citenamefont {Boixo}, \citenamefont {Neven},\ and\ \citenamefont
  {McClean}}]{Huang2021}%
  \BibitemOpen
  \bibfield  {author} {\bibinfo {author} {\bibfnamefont {H.-Y.}\ \bibnamefont
  {Huang}}, \bibinfo {author} {\bibfnamefont {M.}~\bibnamefont {Broughton}},
  \bibinfo {author} {\bibfnamefont {M.}~\bibnamefont {Mohseni}}, \bibinfo
  {author} {\bibfnamefont {R.}~\bibnamefont {Babbush}}, \bibinfo {author}
  {\bibfnamefont {S.}~\bibnamefont {Boixo}}, \bibinfo {author} {\bibfnamefont
  {H.}~\bibnamefont {Neven}},\ and\ \bibinfo {author} {\bibfnamefont {J.~R.}\
  \bibnamefont {McClean}},\ }\bibfield  {title} {\bibinfo {title} {Power of
  data in quantum machine learning},\ }\href
  {https://doi.org/10.1038/s41467-021-22539-9} {\bibfield  {journal} {\bibinfo
  {journal} {Nature Communications}\ }\textbf {\bibinfo {volume} {12}},\
  \bibinfo {pages} {2631} (\bibinfo {year} {2021})}\BibitemShut {NoStop}%
\bibitem [{\citenamefont {Lorena}\ \emph {et~al.}(2019)\citenamefont {Lorena},
  \citenamefont {Garcia}, \citenamefont {Lehmann}, \citenamefont {Souto},\ and\
  \citenamefont {Ho}}]{10.1145/3347711}%
  \BibitemOpen
  \bibfield  {author} {\bibinfo {author} {\bibfnamefont {A.~C.}\ \bibnamefont
  {Lorena}}, \bibinfo {author} {\bibfnamefont {L.~P.~F.}\ \bibnamefont
  {Garcia}}, \bibinfo {author} {\bibfnamefont {J.}~\bibnamefont {Lehmann}},
  \bibinfo {author} {\bibfnamefont {M.~C.~P.}\ \bibnamefont {Souto}},\ and\
  \bibinfo {author} {\bibfnamefont {T.~K.}\ \bibnamefont {Ho}},\ }\bibfield
  {title} {\bibinfo {title} {How complex is your classification problem? a
  survey on measuring classification complexity},\ }\bibfield  {journal}
  {\bibinfo  {journal} {ACM Comput. Surv.}\ }\textbf {\bibinfo {volume} {52}},\
  \href {https://doi.org/10.1145/3347711} {10.1145/3347711} (\bibinfo {year}
  {2019})\BibitemShut {NoStop}%
\bibitem [{\citenamefont {Ho}\ and\ \citenamefont {Basu}(2002)}]{990132}%
  \BibitemOpen
  \bibfield  {author} {\bibinfo {author} {\bibfnamefont {T.~K.}\ \bibnamefont
  {Ho}}\ and\ \bibinfo {author} {\bibfnamefont {M.}~\bibnamefont {Basu}},\
  }\bibfield  {title} {\bibinfo {title} {Complexity measures of supervised
  classification problems},\ }\href {https://doi.org/10.1109/34.990132}
  {\bibfield  {journal} {\bibinfo  {journal} {IEEE Transactions on Pattern
  Analysis and Machine Intelligence}\ }\textbf {\bibinfo {volume} {24}},\
  \bibinfo {pages} {289} (\bibinfo {year} {2002})}\BibitemShut {NoStop}%
\bibitem [{\citenamefont {Sotoca}\ \emph {et~al.}(2005)\citenamefont {Sotoca},
  \citenamefont {S{\'a}nchez},\ and\ \citenamefont {Mollineda}}]{22}%
  \BibitemOpen
  \bibfield  {author} {\bibinfo {author} {\bibfnamefont {J.~M.}\ \bibnamefont
  {Sotoca}}, \bibinfo {author} {\bibfnamefont {J.~S.}\ \bibnamefont
  {S{\'a}nchez}},\ and\ \bibinfo {author} {\bibfnamefont {R.~A.}\ \bibnamefont
  {Mollineda}},\ }\bibfield  {title} {\bibinfo {title} {A review of data
  complexity measures and their applicability to pattern classification
  problems},\ }\href@noop {} {\bibfield  {journal} {\bibinfo  {journal} {Actas
  del III Taller Nacional de Mineria de Datos y Aprendizaje}\ }\textbf
  {\bibinfo {volume} {1}},\ \bibinfo {pages} {18} (\bibinfo {year}
  {2005})}\BibitemShut {NoStop}%
\bibitem [{\citenamefont {Nguyen}\ \emph {et~al.}(2022)\citenamefont {Nguyen},
  \citenamefont {Ho}, \citenamefont {Nguyen~Tran},\ and\ \citenamefont
  {Nguyen}}]{Nguyen_2022}%
  \BibitemOpen
  \bibfield  {author} {\bibinfo {author} {\bibfnamefont {Q.~C.}\ \bibnamefont
  {Nguyen}}, \bibinfo {author} {\bibfnamefont {L.~B.}\ \bibnamefont {Ho}},
  \bibinfo {author} {\bibfnamefont {L.}~\bibnamefont {Nguyen~Tran}},\ and\
  \bibinfo {author} {\bibfnamefont {H.~Q.}\ \bibnamefont {Nguyen}},\ }\bibfield
   {title} {\bibinfo {title} {Qsun: an open-source platform towards practical
  quantum machine learning applications},\ }\href
  {https://doi.org/10.1088/2632-2153/ac5997} {\bibfield  {journal} {\bibinfo
  {journal} {Machine Learning: Science and Technology}\ }\textbf {\bibinfo
  {volume} {3}},\ \bibinfo {pages} {015034} (\bibinfo {year}
  {2022})}\BibitemShut {NoStop}%
\bibitem [{\citenamefont {Komorniczak}\ and\ \citenamefont
  {Ksieniewicz}(2023)}]{komorniczak2023problexity}%
  \BibitemOpen
  \bibfield  {author} {\bibinfo {author} {\bibfnamefont {J.}~\bibnamefont
  {Komorniczak}}\ and\ \bibinfo {author} {\bibfnamefont {P.}~\bibnamefont
  {Ksieniewicz}},\ }\bibfield  {title} {\bibinfo {title} {{problexity---An
  open-source Python library for supervised learning problem complexity
  assessment}},\ }\href@noop {} {\bibfield  {journal} {\bibinfo  {journal}
  {Neurocomputing}\ }\textbf {\bibinfo {volume} {521}},\ \bibinfo {pages} {126}
  (\bibinfo {year} {2023})}\BibitemShut {NoStop}%
\bibitem [{\citenamefont {Cortes}\ and\ \citenamefont
  {Vapnik}(1995)}]{Cortes1995}%
  \BibitemOpen
  \bibfield  {author} {\bibinfo {author} {\bibfnamefont {C.}~\bibnamefont
  {Cortes}}\ and\ \bibinfo {author} {\bibfnamefont {V.}~\bibnamefont
  {Vapnik}},\ }\bibfield  {title} {\bibinfo {title} {Support-vector networks},\
  }\href {https://doi.org/10.1007/BF00994018} {\bibfield  {journal} {\bibinfo
  {journal} {Machine Learning}\ }\textbf {\bibinfo {volume} {20}},\ \bibinfo
  {pages} {273} (\bibinfo {year} {1995})}\BibitemShut {NoStop}%
\bibitem [{\citenamefont {Boughorbel}\ \emph {et~al.}(2017)\citenamefont
  {Boughorbel}, \citenamefont {Jarray},\ and\ \citenamefont
  {El-Anbari}}]{10.1371/journal.pone.0177678}%
  \BibitemOpen
  \bibfield  {author} {\bibinfo {author} {\bibfnamefont {S.}~\bibnamefont
  {Boughorbel}}, \bibinfo {author} {\bibfnamefont {F.}~\bibnamefont {Jarray}},\
  and\ \bibinfo {author} {\bibfnamefont {M.}~\bibnamefont {El-Anbari}},\
  }\bibfield  {title} {\bibinfo {title} {Optimal classifier for imbalanced data
  using matthews correlation coefficient metric},\ }\href
  {https://doi.org/10.1371/journal.pone.0177678} {\bibfield  {journal}
  {\bibinfo  {journal} {PLOS ONE}\ }\textbf {\bibinfo {volume} {12}},\ \bibinfo
  {pages} {1} (\bibinfo {year} {2017})}\BibitemShut {NoStop}%
\bibitem [{\citenamefont {Saunders}\ \emph {et~al.}(1998)\citenamefont
  {Saunders}, \citenamefont {Gammerman},\ and\ \citenamefont
  {Vovk}}]{10.5555/645527.657464}%
  \BibitemOpen
  \bibfield  {author} {\bibinfo {author} {\bibfnamefont {C.}~\bibnamefont
  {Saunders}}, \bibinfo {author} {\bibfnamefont {A.}~\bibnamefont
  {Gammerman}},\ and\ \bibinfo {author} {\bibfnamefont {V.}~\bibnamefont
  {Vovk}},\ }\bibfield  {title} {\bibinfo {title} {Ridge regression learning
  algorithm in dual variables},\ }in\ \href
  {https://doi.org/10.5555/645527.657464} {\emph {\bibinfo {booktitle}
  {Proceedings of the Fifteenth International Conference on Machine
  Learning}}},\ \bibinfo {series and number} {ICML '98}\ (\bibinfo  {publisher}
  {Morgan Kaufmann Publishers Inc.},\ \bibinfo {address} {San Francisco, CA,
  USA},\ \bibinfo {year} {1998})\ p.\ \bibinfo {pages} {515–521}\BibitemShut
  {NoStop}%
\bibitem [{\citenamefont {Kraskov}\ \emph {et~al.}(2004)\citenamefont
  {Kraskov}, \citenamefont {St\"ogbauer},\ and\ \citenamefont
  {Grassberger}}]{PhysRevE.69.066138}%
  \BibitemOpen
  \bibfield  {author} {\bibinfo {author} {\bibfnamefont {A.}~\bibnamefont
  {Kraskov}}, \bibinfo {author} {\bibfnamefont {H.}~\bibnamefont
  {St\"ogbauer}},\ and\ \bibinfo {author} {\bibfnamefont {P.}~\bibnamefont
  {Grassberger}},\ }\bibfield  {title} {\bibinfo {title} {Estimating mutual
  information},\ }\href {https://doi.org/10.1103/PhysRevE.69.066138} {\bibfield
   {journal} {\bibinfo  {journal} {Phys. Rev. E}\ }\textbf {\bibinfo {volume}
  {69}},\ \bibinfo {pages} {066138} (\bibinfo {year} {2004})}\BibitemShut
  {NoStop}%
\bibitem [{\citenamefont {Ross}(2014)}]{10.1371/journal.pone.0087357}%
  \BibitemOpen
  \bibfield  {author} {\bibinfo {author} {\bibfnamefont {B.~C.}\ \bibnamefont
  {Ross}},\ }\bibfield  {title} {\bibinfo {title} {Mutual information between
  discrete and continuous data sets},\ }\href
  {https://doi.org/10.1371/journal.pone.0087357} {\bibfield  {journal}
  {\bibinfo  {journal} {PLOS ONE}\ }\textbf {\bibinfo {volume} {9}},\ \bibinfo
  {pages} {1} (\bibinfo {year} {2014})}\BibitemShut {NoStop}%
\bibitem [{\citenamefont {Kohavi}(1995)}]{1643031.1643047}%
  \BibitemOpen
  \bibfield  {author} {\bibinfo {author} {\bibfnamefont {R.}~\bibnamefont
  {Kohavi}},\ }\bibfield  {title} {\bibinfo {title} {A study of
  cross-validation and bootstrap for accuracy estimation and model selection},\
  }in\ \href@noop {} {\emph {\bibinfo {booktitle} {Proceedings of the 14th
  International Joint Conference on Artificial Intelligence - Volume 2}}},\
  \bibinfo {series and number} {IJCAI'95}\ (\bibinfo  {publisher} {Morgan
  Kaufmann Publishers Inc.},\ \bibinfo {address} {San Francisco, CA, USA},\
  \bibinfo {year} {1995})\ p.\ \bibinfo {pages} {1137–1143}\BibitemShut
  {NoStop}%
\bibitem [{\citenamefont {Savage}(1951)}]{Savage1951}%
  \BibitemOpen
  \bibfield  {author} {\bibinfo {author} {\bibfnamefont {L.~J.}\ \bibnamefont
  {Savage}},\ }\bibfield  {title} {\bibinfo {title} {The theory of statistical
  decision},\ }\href {https://doi.org/10.2307/2280094} {\bibfield  {journal}
  {\bibinfo  {journal} {Journal of the American Statistical Association}\
  }\textbf {\bibinfo {volume} {46}},\ \bibinfo {pages} {55} (\bibinfo {year}
  {1951})},\ \bibinfo {note} {full publication date: Mar., 1951}\BibitemShut
  {NoStop}%
\bibitem [{\citenamefont {Dem{\v{s}}ar}(2006)}]{JMLR:v7:demsar06a}%
  \BibitemOpen
  \bibfield  {author} {\bibinfo {author} {\bibfnamefont {J.}~\bibnamefont
  {Dem{\v{s}}ar}},\ }\bibfield  {title} {\bibinfo {title} {Statistical
  comparisons of classifiers over multiple data sets},\ }\href
  {http://jmlr.org/papers/v7/demsar06a.html} {\bibfield  {journal} {\bibinfo
  {journal} {Journal of Machine Learning Research}\ }\textbf {\bibinfo {volume}
  {7}},\ \bibinfo {pages} {1} (\bibinfo {year} {2006})}\BibitemShut {NoStop}%
\bibitem [{\citenamefont {Lyon}(2015)}]{htru2_372}%
  \BibitemOpen
  \bibfield  {author} {\bibinfo {author} {\bibfnamefont {R.}~\bibnamefont
  {Lyon}},\ }\href@noop {} {\bibinfo {title} {{HTRU2}}},\ \bibinfo
  {howpublished} {UCI Machine Learning Repository} (\bibinfo {year} {2015}),\
  \bibinfo {note} {{DOI}: https://doi.org/10.24432/C5DK6R}\BibitemShut
  {NoStop}%
\bibitem [{\citenamefont {Lorena}\ \emph {et~al.}(2018)\citenamefont {Lorena},
  \citenamefont {Maciel}, \citenamefont {de~Miranda}, \citenamefont {Costa},\
  and\ \citenamefont {Prud{\^e}ncio}}]{Lorena2018}%
  \BibitemOpen
  \bibfield  {author} {\bibinfo {author} {\bibfnamefont {A.~C.}\ \bibnamefont
  {Lorena}}, \bibinfo {author} {\bibfnamefont {A.~I.}\ \bibnamefont {Maciel}},
  \bibinfo {author} {\bibfnamefont {P.~B.~C.}\ \bibnamefont {de~Miranda}},
  \bibinfo {author} {\bibfnamefont {I.~G.}\ \bibnamefont {Costa}},\ and\
  \bibinfo {author} {\bibfnamefont {R.~B.~C.}\ \bibnamefont {Prud{\^e}ncio}},\
  }\bibfield  {title} {\bibinfo {title} {Data complexity meta-features for
  regression problems},\ }\href {https://doi.org/10.1007/s10994-017-5681-1}
  {\bibfield  {journal} {\bibinfo  {journal} {Machine Learning}\ }\textbf
  {\bibinfo {volume} {107}},\ \bibinfo {pages} {209} (\bibinfo {year}
  {2018})}\BibitemShut {NoStop}%
\end{thebibliography}%

\end{document}